\documentclass[5p]{elsarticle} % seleccionar: preprint, review, 1p, 3p, 5p

\usepackage{mathtools}
\usepackage{subcaption}
\usepackage{longtable}
\journal{ }

\usepackage{placeins}
\usepackage{eurosym}
\usepackage{threeparttable} % allow the use of footnote within tables

\usepackage{url}

\usepackage{hyperref}
\hypersetup{colorlinks,allcolors=black}

\usepackage{lineno}
\modulolinenumbers[5]

\usepackage{lineno,hyperref}
\modulolinenumbers[1]
\usepackage{amsmath}
\usepackage{siunitx}
\usepackage{eurosym}
\biboptions{numbers,sort&compress}
\usepackage[europeanresistors,americaninductors]{circuitikz}
\usepackage{adjustbox}
\usepackage{xspace}
\usepackage{caption}
\usepackage{booktabs}
\usepackage{tabularx}
\usepackage{multicol}
\usepackage{float}
\usepackage{graphicx,dblfloatfix}
\usepackage{csvsimple}
\begin{document}
\begin{frontmatter}

\title{Comparative analysis of PV configurations for agrivoltaic systems in Europe}

\author[mymainaddress]{Kamran Ali Khan Niazi}

\author[mymainaddress,CORC]{Marta Victoria}

% \corref{mycorrespondingauthor}}
%\ead{mvp@mpe.au.dk}
\address[mymainaddress]{Department of Mechanical and Production Engineering, iClimate, Aarhus University, Katrinebjergvej 89G-F, 
8200 Aarhus N, Denmark}
\address[CORC]{Novo Nordisk Foundation CO2 Research Center, Gustav Wieds Vej 10, Aarhus 8000, Denmark}

\begin{abstract}
Agrivoltaics (APV) is the dual use of land by combining agricultural crop production and photovoltaic (PV) systems. In this work, we have analyzed three different APV configurations:  static with optimal tilt, vertically-mounted bifacial, and single-axis horizontal tracking. A model is developed to calculate the shadowing losses on the PV panels along with the reduced solar irradiation reaching the area under them for different PV capacity densities. First, we investigate the trade-offs using a location in Denmark as a case study and second, we extrapolate the analysis to the rest of Europe. We find that the vertical and single-axis tracking produce more uniform irradiance on the ground, and a capacity density of around 30 W/m$^2$ is suitable for APV systems. Based on our model and a 100 m-resolution land cover database, we calculate the potential for APV in every NUTS-2 region within the European Union (EU). The potential for APV is enormous as the electricity generated by APV systems could produce 28 times the current electricity demand in Europe. Overall, the potential capacity for APV in Europe is 51 TW, which would result in an electricity yield of 71500 TWh/year.
\end{abstract}
\end{frontmatter}

%\linenumbers
\section{Introduction} % Marta (500 words)

%Task 5.1: Identification potential EU-sites for agrivoltaics (AU)
%Task 5.1: Estimation of the potential for agrivoltaics in different European regions (including NUTS 1 and 2) based on land-use databases (with a spatial resolution of approximately 100 m) and solar radiation database from satellite measurements. An agrivoltaics potential atlas for Europe will be produced and it will be validated with the data from demonstration sites in the project. 
%D5.1: Journal publication describing the potential of agrivoltaics and open database with the estimated potentials per region (M24)

\begin{figure}[!h]
\centering
\begin{subfigure}{0.47\textwidth}
\includegraphics[width=\linewidth]{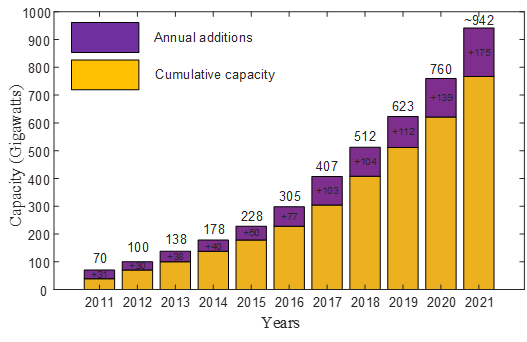}
\caption{}
\label{fig:a}
\end{subfigure}
\begin{subfigure}{0.47\textwidth}
\includegraphics[width=\linewidth]{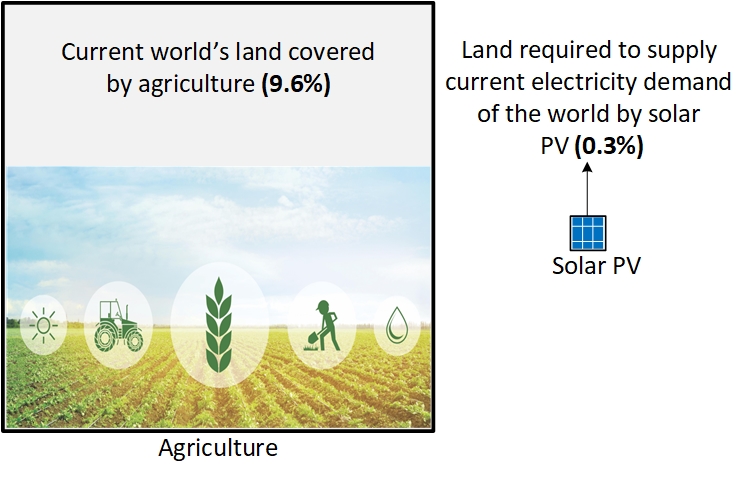}
\caption{}
\label{fig:b}
\end{subfigure}
\caption{(a) Global solar photovoltaic (PV) installed capacity over the years \cite{winter2022renewables} (b) current land covered by agriculture vs maximum land that would be required by solar PV to supply world electricity demand today \cite{Victoria_2019}.} \label{global_production} 
\end{figure}

%A typical research article will be 5,000 words of text (excluding methods and references) with up to 7 display items. 
Solar photovoltaics (PV) has shown a fast deployment in the last decade and its cumulative global capacity attained 942 GW in 2021 \cite{winter2022renewables}, see Fig. \ref{global_production}(a). From a global perspective, land used by solar PV should not represent a problem. For instance, the current global electricity consumption could be supplied by solar PV covering only 0.3\% of the land area of the world \cite{victoria2021solar}. However, local competition for land uses can be problematic and the concentrated deployment of solar PV plants can trigger social acceptance issues. At the same time, the preservation of agricultural land, sustainable increase in crop yield, and adaptation to climate change are the most relevant challenges for global agriculture. \cite{rehbein2020renewable, alam2022techno}. In some cases, the combination with solar PV systems can be a strategy to address those challenges and can provide mutual benefits for the PV systems and the crops. Today, about 9.6\% of the world’s land is used for agriculture. Combining with PV systems part of that land will untap a large potential for sustainable generation of electricity, as depicted in Fig. \ref{global_production}(b).  Additionally, this sustainable food-energy cooperation could also add biological reservoirs and help in preserving the terrestrial ecosystems and biodiversity \cite{Horowitz_2020}.

APV is defined as the simultaneous use of land for agriculture and PV systems \cite{andrew2021herbage,trommsdorff2020agrivoltaics,al2022review}. Synergies can enable both the crops and the PV modules to benefit from this integration. In dry climates, the shadow cast by PV modules could reduce the irrigation needs by up to 20\% due to an altered micro-climate below them \cite{elamri2018water, Adeh_2019}. In addition, solar panels could also be used to collect rainwater, which can then be used for irrigation especially for dry lands \cite{Barron-Gafford_2019}, although this depends on the mounting structure. Another possible benefit for crops is that they could be protected from weather influences like heavy rainfall, hail, or wind by the PV modules themselves or by foil tunnels, which could use the mounting structure of the modules. A potential benefit for the PV system would be an increased performance ratio due to improved convective cooling \cite{trommsdorff2020agrivoltaics}. 

This work aims to investigate the potential of APV installations across Europe. For this purpose, three different configurations for PV systems are investigated under different spacing between the rows and various heights. As expected, the amount of electricity generated is highest for the horizontal single-axis tracking setup, as compared to the optimal tilt installation and the vertical bifacial setup. The losses due to PV modules self-shadowing depend on the configuration and the capacity density. 

Additionally, the impact of the PV installation on the field is also investigated to determine the potential for agricultural use. The irradiance distribution on the ground shows relatively even irradiance distribution for vertical bifacial and single axis-tracking setup. The tilted setup causes a distinct pattern of stripes with substantial shadowing. 

These analyses are extended to determine the potential for APV along with the eligible area in European countries. Lastly, the electricity yield over the year is determined for APV configurations by considering the specific land types and a capacity density of 30 W/m$^2$. 

Overall, this paper is divided into various sections. An introduction has been provided in Section 1 that leads to the investigated APV configurations, which are presented in Section 2. Section 3 describes the mathematical modeling and Section 4 consists of results and discussion. Finally, the conclusion of the work is drawn in Section 5.

%%%%%%%%%%%%%%%%%%%%%%%%%%%%%%%%%%%%%%%%%%%%%%%%%%%%%%%%%%%%%%%%%%%%%%%%%%%%%%%%%%%%%%%%%%%%%%%%% 
\section{Investigated Agrivoltaic Configurations} \label{sec_methods} % description of systems
% First, a detailed study of the electricity output and shadow cast on the ground is carried out in order to analyse performance indicators for different configurations in an APV system in Foulum, Denmark (latitude: 56.49$^{\circ}$, longitude: 9.57$^{\circ}$). 
Three APV configurations are investigated in this work across Europe: optimal tilted, vertical bifacial, and horizontal single-axis tracking. A detailed description of their structure and distinguishing features is presented below.

\subsection{Optimal tilted PV system}
 A south-facing monofacial fixed-tilt PV system is considered in this configuration. The optimal tilt angle of the system, varies depending on the latitude in order to maximize the annual electricity generation, according to \cite{PVGIS}. A schematic of the optimal tilted installation, which will be referred to as tilted configuration in the following, is shown in Fig. \ref{configurations}(a).
 
 \subsection{Horizontal single-axis tracking PV system}
The tilt angle of monofacial PV panels, which are mounted on a north-south axis varies continuously throughout the day in this configuration. The PV panels face east in the morning, are horizontal at noon, and face west in the evening. A schematic of the horizontal single-axis tracking setup for different times of the day is shown in Fig. \ref{configurations}(b) This configuration is henceforth referred to as single-axis tracking.

\subsection{Vertical bifacial PV system}
\begin{figure}[!]
\centering
\begin{subfigure}{0.5\textwidth}
\includegraphics[width=\linewidth]{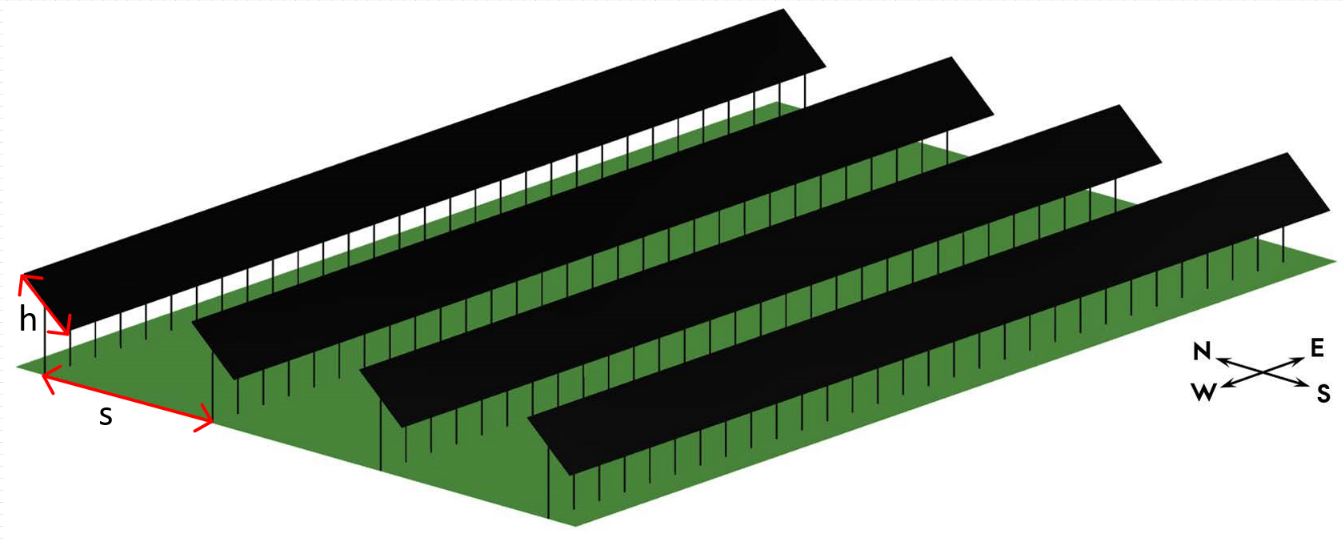}
\caption{}
\label{fig:subim1}
\end{subfigure}
\begin{subfigure}{0.5\textwidth}
\includegraphics[width=\linewidth]{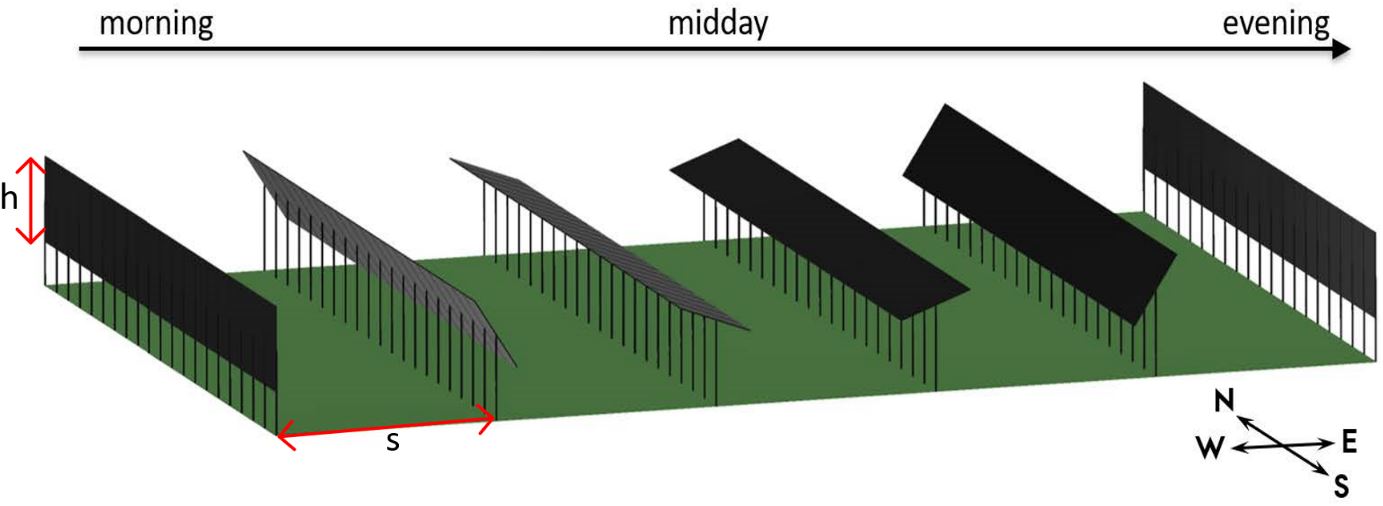}
\caption{}
\label{fig:subim1}
\end{subfigure}
\begin{subfigure}{0.5\textwidth}
\includegraphics[width=\linewidth]{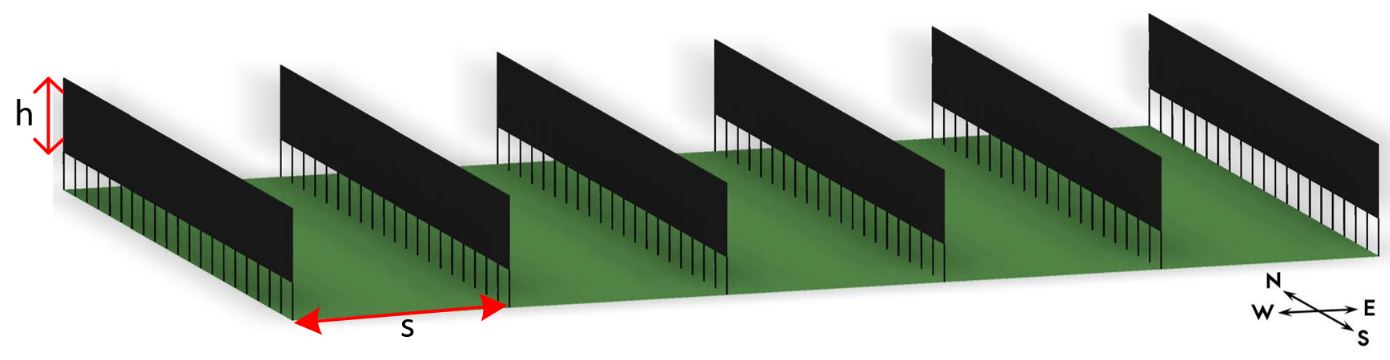}
\caption{}
\label{fig:subim1}
\end{subfigure}
\caption{Solar PV configurations analyzed in this work: (a) static with optimal tilt, (b) single-axis horizontal tracking, and (c) vertical mounted bifacial. The parameters inter-row spacing $s$ and height $h$ are shown in the figure.}

\label{configurations} 
\end{figure}

In a vertical bifacial setup, PV modules are placed vertically on a north-south line. Therefore, one side of the PV modules faces east while the other side faces west. A schematic of this vertical bifacial setup is shown in Fig. \ref{configurations}(c) and this configuration is referred to as vertical bifacial. A bifaciality factor equal to 0.8 is assumed.

%%%%%%%%%%%%%%%%%%%%%%%%%%%%%%%%%%%%%%%%%%%%%%%%%%%%%%%%%%%%%%%%%%%%%%%%%%%%%%%%%%%%%%%%%%%%%%%%% 
\section{Methods/Experimental Procedures} \label{sec_methods} %(1000 words + Suplemental Materials)
The three setups are investigated on a reference field with a size of 100m x 100m when analyzing their electricity output and on a 50m x 50m field to determine shadowing effects on the underlying ground. These field sizes were chosen to minimize errors due to border effects, while at the same time keeping the computational effort reasonable. The solar PV rows are installed at a fixed distance above the ground, which is 2m for the tilted setup while 1m for the axis-tracking and vertical bifacial setup. This is because the tilted setup needs a higher elevation for harvesting. In all cases, the distance above the ground is measured between the ground and the lowest point of the solar panel. The distance from the ground does not influence electricity production, as the shadows cast from one row of solar panels onto another are not affected. Conversely, the shadow distribution on the ground is affected by this parameter. Two other variables whose impact on the system is investigated in conjunction with the three different installation types are the spacing between the rows $s$ and the height $h$ of the PV modules, as depicted in Fig. \ref{configurations}.

In our analysis, the height can take a value of 1, 2, or 3m (in practice this can be achieved by stacking several PV panels). The inter-row spacing can take a value of 3, 4.5, 6, 7.5, 9, and 12m. These two parameters enable to study the differences between dense and sparse installations.

\subsection{Modelling solar PV generation}
To compare different configurations, it is assumed that all setups use solar panels with similar electrical properties, both the monofacial and vertical bifacial panels. The reference solar panel is the N-type bifacial high-efficiency mono silicon double glass panel produced by Jolywood, which has a bifaciality factor of 80\% \cite{jolywood}. Each of the panels contains 72 individual series-connected solar cells, which are divided into 3 equal blocks with bypass diodes.

% % \begin{table}[!ht]
% \begin{table}[!ht]
%     \centering
%     \caption{Installation positioning parameters.}
%     \begin{tabular}{|l|l|l|l|}
%     \hline
%         Spacing, s (m) & 3 & 4.5 & 6 \\ 
%         & 7.5 & 9 & 12 \\ \hline
%         Height, h (m) & 1 & 2 & 3 \\ \hline
%     \end{tabular}
% \end{table}

\subsubsection{Solar irradiance reaching the solar PV panels}

The global $G(0)$ irradiance is composed of direct $B(0)$, diffuse $D(0)$, and albedo $R(0)$ irradiance, which are obtained from PVGIS \cite{PVGIS}. The direct irradiance $B(\beta, \alpha)$ can be calculated by using Eq \ref{eqn_direct}.

\begin{equation} \label{eqn_direct}
	B(\beta,\alpha) = \frac{B(0) max(0,cos\theta_s)}{sin\gamma_s} 
\end{equation} here $\beta$\ is tilt angle, $\alpha$ is the angle of orientation of the PV module, $\gamma_s$ is solar altitude, and $\theta_s$ is the angle of incidence, which is defined as the angle between the surface normal and the vector of the radiation coming directly from the sun \cite{Victoria_2019}.

The diffuse irradiance $D$ is assumed to be composed of two parts, diffuse circumsolar irradiance $D^{circ}$ and diffuse isotropic irradiance $D^{iso}$, Eq \eqref{eqn_D} \cite{Victoria_2019}. This approach uses the anisotropic model by Hay and Davies \cite{hay1985estimating}. The anisotropy index $k_1$ is determined, as the ratio of the horizontal direct irradiance on the ground $B(0)$ and at the top of the atmosphere $B_0(0)$ according to Eq \eqref{eqn_anisotropy_index}.

\begin{equation} \label{eqn_D}
	D = {D^{circ}+D^{iso}} 
\end{equation}
	
\begin{equation} \label{eqn_anisotropy_index}
	k_1 = \frac{B(0)} {B_0(0)}
	\end{equation}
	
To calculate the diffuse circumsolar $D^{circ}$ and diffuse isotopic irradiance $D^{iso}$, Eqs \eqref{eqn_Dcirc} and \eqref{eqn_Diso} are used, respectively. Here $D(0)$ is the horizontal diffuse irradiance on the ground and $k_{hori}$ is the horizon brightening effect\cite{reindl1990evaluation} given in Eq \eqref{eqn_hori}.

\begin{equation} \label{eqn_Dcirc}
	D^{circ}(\beta, \alpha) = k_1.\frac{D(0) max(0, cos\theta_s)} {sin\gamma_s}
\end{equation}

\begin{equation} \label{eqn_Diso}
	D^{iso}(\beta, \alpha) = k_{hori}(1-k_1)D(0)\frac{1+cos\beta} {2}
\end{equation} 

\begin{equation} \label{eqn_hori}
	k_{hori}=1 + \sqrt{1-F}sin^3(\frac{\gamma_s}{2})
\end{equation} here $F$ is the diffuse fraction of the global irradiance.

Lastly, the albedo irradiance $R(\beta,\alpha)$ can be calculated as
\begin{equation} \label{eqn_albedo}
	R(\beta,\alpha)=\rho G(0) \frac{1-cos\beta}{2}
\end{equation} where $\rho$ is the reflectivity of the ground \cite{Victoria_2019}.

It is assumed that only the direct and diffuse circumsolar components are blocked by objects and therefore create shadows since they have a clearly defined directional vector. In contrast to that, the diffuse isotropic and reflected irradiance reaching the investigated surfaces are not impacted by shadows.

The global irradiance on the surface of the PV module is
\begin{equation} \label{eqn_global_irr}
	G(\beta,\alpha)=B(\beta,\alpha)+D(\beta,\alpha)+R(\beta,\alpha)
\end{equation}

\subsubsection{Solar electricity production}
One of the main aspects used to compare the different setups is the power output, which can be calculated using Eq \ref{eqn_elec_power}.

\begin{equation} \label{eqn_elec_power}
	P = P_{STC}.\eta.\eta_{sys}\frac{(B+D^{circ})(1-F_{ES})(1-AL)+D^{iso}+R}{G_{STC}}
\end{equation} Here $\eta_{sys}$ corresponds to the overall system losses, while the combined effects due to
cell temperature and low irradiance are represented by $\eta$. The losses impacting the direct irradiance $B$ and the diffuse circumsolar irradiance $D_{circ}$ are shadow losses and losses due to the reflection of light at the enterence of the PV module. These are incorporated using the shading factor $F_{ES}$ and the angular loss factor $AL$. It is assumed that the diffuse isotropic $D_{iso}$ and albedo
irradiance $R$ are not affected by those two factors.

The power losses due to shadowing are included by using the model from \cite{martinez2010experimental}, which is considering the shadowed area of the PV panels along with the circuit structure. This is necessary since any type of shadowing will result in a reduction in electricity output for the entire panel. Using this model, the effective shading factor $F_{ES}$ of a solar panel in Eq \ref{eqn_elec_power} is calculated according to Eq \ref{eqn_FES}.

\begin{equation} \label{eqn_FES}
	(1-F{_{ES}}) = (1-F{_{GS}})({1-\frac{N{_{SB}}}{N{_{TB}}+1})}
\end{equation} here $F_{GS}$ is defined as the geometric shading factor, which is the fraction of the total area being shaded. $N_{TB}$ and $N_{SB}$ are the number of total and shaded blocks, respectively. A block is defined as a string of series-connected solar cells protected by a bypass diode. The PV panels in our model have $N_{TB}$=3. Moreover, the solar panels are assumed to be in landscape orientation to minimize the shadow losses.

Angular losses in Eq \ref{eqn_elec_power} are modelled using Eq \ref{eqn_angular}.

\begin{equation} \label{eqn_angular}
	AL(\theta_s) =  1-[\frac{1-exp(\frac{-cos\theta_s}{\alpha_r})}{1-exp(\frac{-1}{\alpha_r})}]
\end{equation} here $\theta_s$ is the angle of incident irradiance and $\alpha_r$ the angular losses coefficient. This coefficient depends on the solar cell type and the amount of soiling present on the panel’s surface. Throughout this study, a factor of $\alpha_r$ = 0.17 is used according to \cite{martin2001calculation}. The chosen $\alpha_r$ value corresponds to a typical value for a silicon solar cell without soiling and losses due to soiling are included in the system losses $\eta_{sys}$.

Ambient temperature $T_{amb}$ and wind speed are obtained from PVGIS. $T_{amb}$ is converted to cell temperature $T_{cell}$ by using Eq \ref{eqn_Tmod}.

\begin{equation} \label{eqn_Tmod}
	T_{cell} =  T_{amb}+(\frac{G(\beta,\alpha)}{U_0+U_1.W_{mod}})
\end{equation} here $T_{amb}$ is the ambient temperature and $G(\beta, \alpha)$ is the incident irradiance on the panel. The coefficients $U_0$ and $U_1$ depend on the type of solar panel. As we are using a glass crystalline
silicon solar panel, these values are $U_0$ = 26.92 W/m$^2$K and $U_1$ = 6.24 W/m$^3s$K. Moreover, $W_{mod}$ is the wind speed at the solar PV panel and it can be calculated by using Eq \ref{eqn_Wmod}.

\begin{equation} \label{eqn_Wmod}
	W_{mod} =  (\frac{h_{mod}}{h_{10}})^2.W_{10}
\end{equation} where $W_{mod}$ and $W_{10}$ are the wind speed at the PV panel and at a height of 10 meters, respectively. Similarly, $h_{mod}$ and $h_{10}$ are the height of the solar panel and 10 meters. For this, the height is chosen to be the center of the panel.

$T_{cell}$ and $W_{10}$ are obtained from PVGIS \cite{PVGIS}. Furthermore, knowing the cell temperature and incident irradiance, the panel efficiency relative to the one at standard test conditions (STC) conditions $\eta$ can be modeled as follows according to \cite{huld2011power}.
\begin{equation} \label{eqn}
\begin{aligned}
	\eta(G',T') & =  1+k_1ln(G')+k_2(ln(G'))^2+k_3T'\\
      & +k_4T'ln(G')+k_5T'(ln(G'))^2+k_6T'^2
 \end{aligned}
\end{equation} where
\begin{equation*}
	G' =  \frac{G}{G_{STC}}
\end{equation*}
\begin{equation*}
	T'=T_{cell}-T_{STC}
\end{equation*} here $G'$ is the normalized irradiance, $T'$ is the temperature difference, and $T_{STC}$ is the cell temperature at standard test conditions (STC). The coefficients $k_1-k_6$ depend on the PV panel type. Here, we have assumed crystalline-silicon PV panel and the values are taken from \cite{huld2011power}.

%Describe the modelling of radiation reaching the solar panels including data source for global radiation and transformation from ground to tilt radiation. Describe modelling of solar electricity production: temperature coefficient, bifaciality factor, model for shadows, bias correction for angular and low irradiance. Appendix: section 2.2, 2.3.1, 2.3.2, 2.3.3, 2.3.4

\subsection{Modelling shadowing on the ground and impact on crops}

Crops use solar radiation to carry out photosynthesis, but this only happens during some periods of the year and only uses part of the solar spectrum. The latter is quantified by photosynthetically active radiation (PAR), which comprises light of wavelengths between 400 and 700 nm. PAR is usually expressed as the number of photons received by a surface during a specific amount of time and its units are $\mu$mol/m$^2$s· The standard solar spectrum AM1.5G contains 430 W/m$^2$ between 400 and 700 nm. Considering a conversion factor of 4.56 $\mu$mol/Ws, AM1.5G is equivalent to 1960 $\mu$mol/m$^2$s. 

In principle, it is possible to shadow the crops without affecting their yield or even improve it if the shadows contribute to avoiding excessive radiation and high temperature. In reality, the impact of shadowing on crops is very dependent on the site and specific crop. Plants are dynamic organisms whose growth can be limited by available radiation but also by the ratio of direct to diffuse illumination, water and nutrients availability, ambient temperature, humidity, and the size of the leaves grown in previous periods, among other things. Hence, it is difficult to provide a general rule and every crop and location needs to be carefully examined to design an APV system. 

\begin{figure}[!]
\centering
\includegraphics[width=\linewidth]{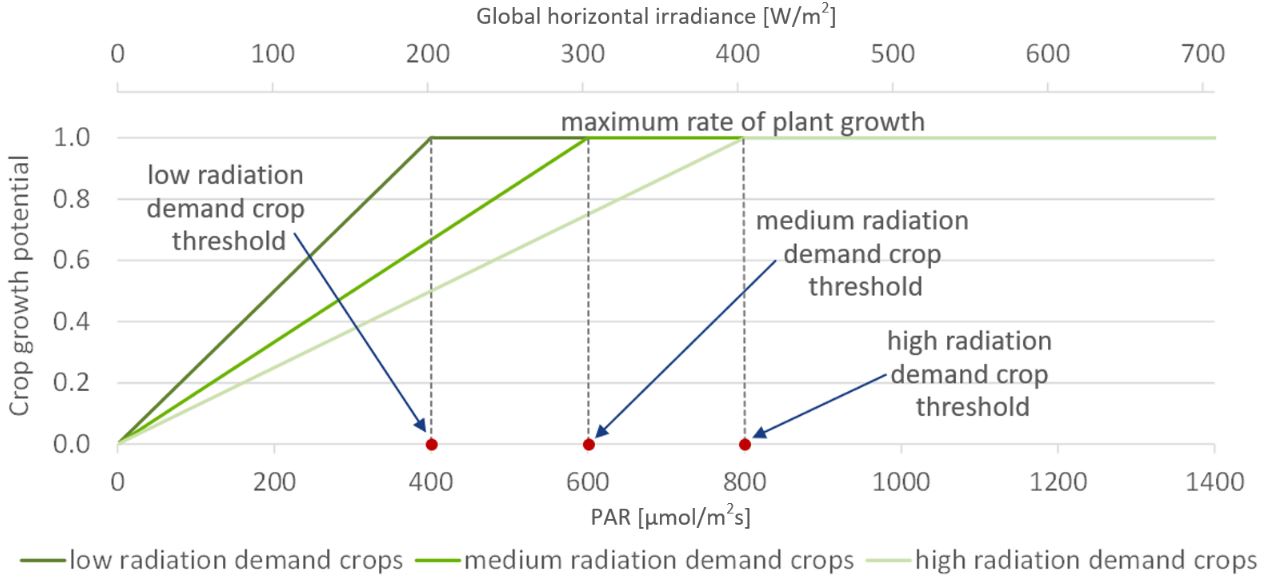}
\caption{ A simplified model for the crop growth potential as a function of the photosynthetically active radiation (PAR) for 3 stylized crops with different saturation thresholds named as low, medium, and high radiation demand plants.}
\label{fig_crops_model} 
\end{figure}

Here, we have a simplified approach in which we assume three categories for crops, those requiring low, medium, and high radiation. We assume that the crops' growth potential is not affected if PAR is above a certain threshold, whose value depends on the type of crop and is shown in Fig. \ref{fig_crops_model}. Then, using the irradiation patterns on the ground (presented later in Fig. \ref{fig_heatmaps}), we calculated for every APV configuration the percentage of the land that could be cultivated without any impact on the crops (because the irradiance remains above the crop threshold).

\subsection{Performance indicators}
The performance of the different APV configurations is evaluated on the basis of the following indicators:
\begin{itemize}
    \item Capacity density (W/m$^2$): It is the average capacity on a reference ground square meter and is calculated by dividing the total installed peak power $C$ by the field area $A_f$.

\begin{equation} \label{eqn_capacity_density}
\begin{aligned}
\textit{Capacity density}=\frac{C}{A_f}
 \end{aligned}
\end{equation}

    \item Electricity yield (kWh/m$^2$): It is calculated by dividing the hourly electricity production $E(h)$ integrated over a whole year by the installed capacity $A_f$.
\begin{equation} \label{eqn_electricity_yield}
\begin{aligned}
\textit{Electricity yield}=\frac{\Sigma_{h}E(h)}{A_f}
 \end{aligned}
\end{equation}
    \item Price-weighted electricity yield (kWh/m$^2$): It is defined as the annual sum of the electricity production every hour $E(h)$ weighted by the spot market electricity price $p(h)$. The result is divided by the field area $A_f$.

\begin{equation} \label{eqn_price_weighted_yield}
\begin{aligned}
\textit{Price-weighted electricity yield } (kWh/m^2)= \\ \frac{\Sigma_{h}E(h).\frac{p(h)}{<p(h)>}}{A_f}
 \end{aligned}
\end{equation} where $<$p(h)$>$ is the average electricity price.
    
    \item Shadow losses (\%): It is defined as the losses in the system due to shadowing and it is calculated by finding the difference between electricity generation with and without shadow effects and then dividing it by the electricity output without shadowing.
    
    \item Specific yield (kWh/kW): It is electricity yield per installed capacity, which can be calculated by dividing the hourly electricity production $E(h)$ integrated over a whole year by the installed capacity $C$.

\begin{equation} \label{eqn_specific_yield}
\begin{aligned}
\textit{Specific yield}=\frac{\Sigma_{h}E(h)}{C}
 \end{aligned}
\end{equation}

\end{itemize}
The results are presented in coming sections by considering these performance indicators.
%\begin{itemize}
%\item Capacity density (W/m$^2$) 
%\item Electricity yield (kWh/m$^2$a) 
%\item Price-Weighted electricity yield (kWh/m$^2$a)
%\item Levelized cost of electricity (LCOE, EURO/MWh)
%%\item Shadow loss (\%)
%item Specific yield (kWh/kW$_p$)

%Add formula
%\end{itemize}

\subsection{Land availability estimation}

To assess the potential of APV systems in Europe, the land feasibility and maximum possible electricity generation are investigated. For this purpose, the suitable available area for APV is determined by performing a land eligibility analysis by using Atlite \cite{hofmann2021atlite}. The selected land types from the Corine Land Cover database \cite{CLC} that are considered suitable for the APV system are given in Table 1.
The selection criteria exclude different types of protected areas, where building projects are prohibited. This includes protected habitats for birds and other wildlife, landscapes and also parks, and natural monuments. For all areas contained in the exclusion criterion, a minimal distance of 100 m is chosen.

\begin{table*}[!ht] \label{table_landtypes}
    \centering
    \caption{Land types considered valid for APV systems in the Corine Land Cover database (100 m space resolution is used)\cite{CLC}.}
    \begin{tabular}{|l|l|}
    \hline
        Considered land types in agricultures areas & Selected areas \\ \hline
        Arable land & Non-irrigated arable land, Permanently irrigated land, \\ & and Rice fields \\ \hline
        Permanent crops & Fruit trees and berry plantations \\ \hline
        Pastures & Pastures, Heterogeneous agricultural areas, Annual crops \\ 
        &  associated with permanent crops, and Complex cultivation patterns \\ \hline
    \end{tabular}
\end{table*}

%%%%%%%%%%%%%%%%%%%%%%%%%%%%%%%%%%%%%%%%%%%%%%%%%%%%%%%%%%%%%%%%%%%%%%%%%%%%%%%%%%%%%%%%%%%%%%%%%%
\section{Results and Discussion} %(3000 words)

\subsection{System evaluation for a Northern location with low irradiance values}

\begin{figure}[!h]
\begin{subfigure}{0.50\textwidth}
\centering
\includegraphics[width=\linewidth]{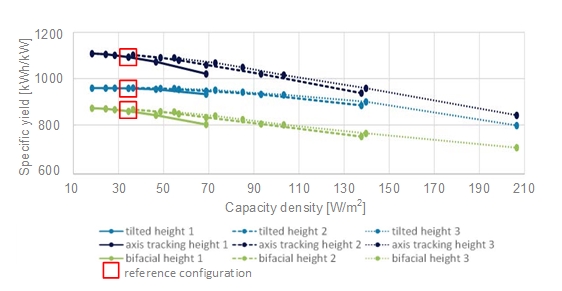}
\caption{}
\label{fig:subim1}
\end{subfigure}
\begin{subfigure}{0.49\textwidth}
\includegraphics[width=\linewidth]{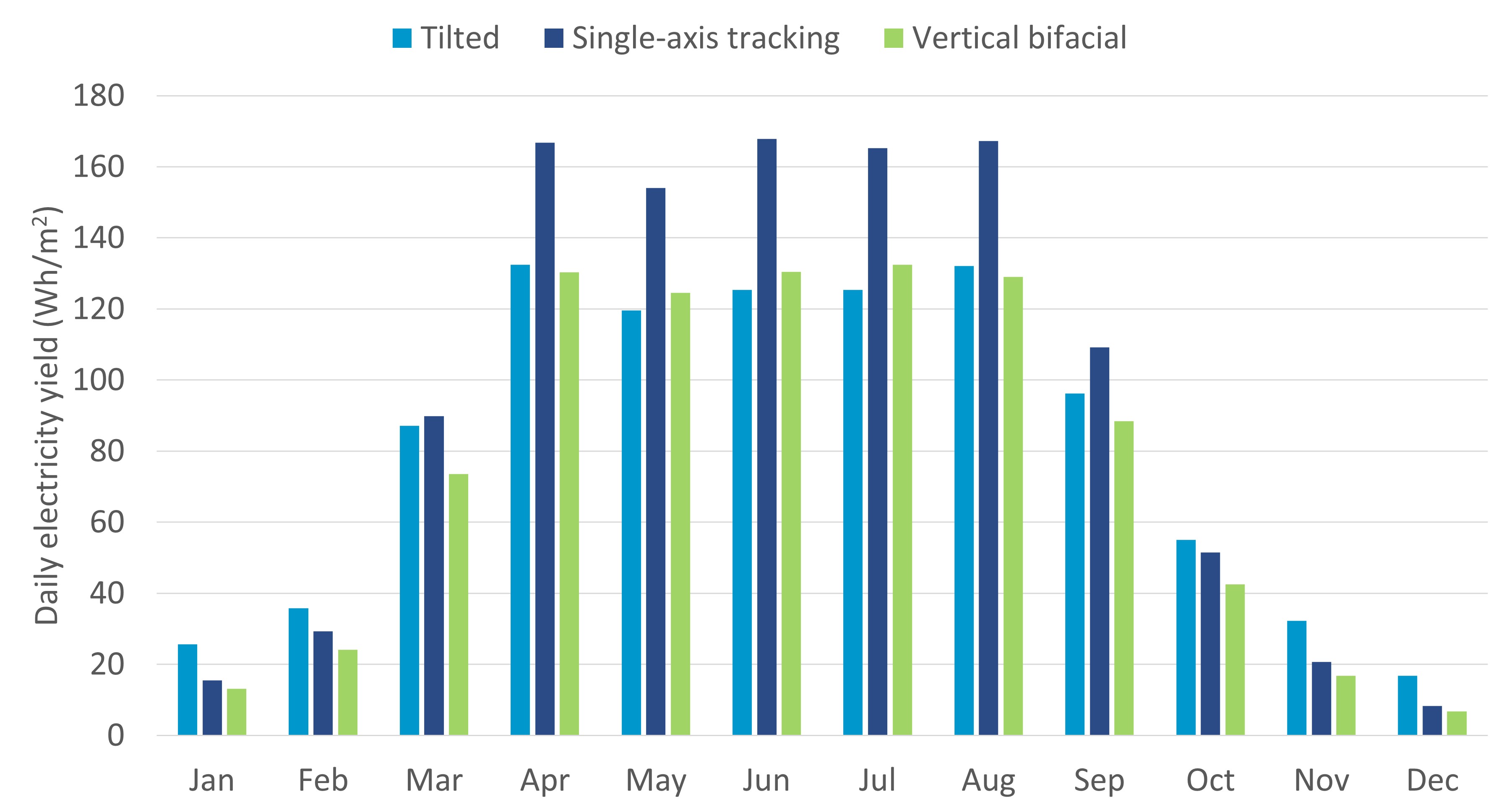}
\caption{}
\label{fig:subim1}
\end{subfigure}
\begin{subfigure}{0.49\textwidth}
\centering
\includegraphics[width=\linewidth]{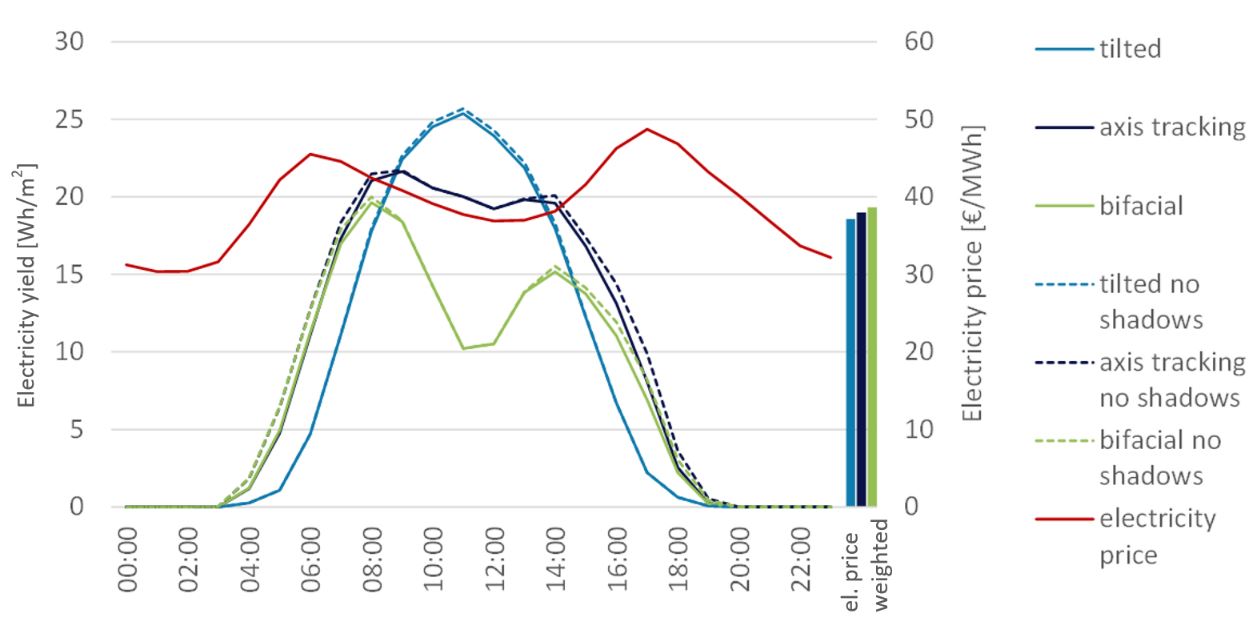}
\caption{}
\label{fig:subim1}
\end{subfigure}

\caption{(a) Specific yield (kWh/kW) as a function of the capacity density for the three configurations under analysis assumed to be installed in Foulum, Denmark (latitude: 56.49$^{\circ}$, longitude: 9.57$^{\circ}$). The inter-row distance and height are defined in Fig. 2, (b) average daily electricity yield for each month for the reference configuration (height= 2 m, spacing= 6 m), and (c) comparison between the electricity generation in Foulum and the electricity price at market DK1 for year 2015 throughout a day for the reference configuration. The bars shows the price-weighted electricity yield calculated with Eq \ref{eqn_price_weighted_yield}.}

\label{fig_electricity_yield} 
\end{figure}

For various PV configurations, the specific yield including shadow losses is shown in Fig. \ref{fig_electricity_yield}(a) for Foulum, Denmark (latitude: 56.49$^{\circ}$, longitude: 9.57$^{\circ}$ ). We have defined a reference setup with s= 6m and h= 2m. As expected, the axis-tracking installation has the highest specific yield, followed by the tilted configuration and at last the vertical bifacial setup. Here, the tilted setup has a nearly constant specific yield until a capacity density of around 70 W/m$^2$, while the other two setups show a decrease at a lower capacity density. For the lowest capacity density considered (15 W/m$^2$) shadow losses are negligible for the three configurations.

Comparing the different heights of one setup type at a specific spacing, it can be seen that the setups with a larger height (displayed with dashed and dotted lines) generate more electricity, as the amount of installed solar PV capacity per square meter of ground is larger. When considering the spacing, the greater the distance between the rows, the lower the number of panels installed in one squared meter of land and thus the lower the amount of electricity generated. However, there is no linear relationship between electricity generation and spacing due to the effect of shadows.

Lets us look now at the temporal generation patterns of the different configurations. The average daily electricity yield for each month is analyzed in Fig. \ref{fig_electricity_yield}(b). As expected for every setup, the electricity production in summer is greater than in winter, due to seasonality. Additionally, it can be seen that in the winter months the tilted system produces more energy than the other two systems, while in the summer months the axis-tracking setup produces the highest amount of energy.

% Overall, the tilted system looses a small amount of energy due to shadow effects in the winter months (see striped parts of the bar in Fig. \ref{fig_electricity_yield}(b)) and nothing in the months from April to September. Similarly, the other two setups loose a higher percentage of the generated electricity due to shadows in the winter months compared to the summer months. However, the absolute loss is higher in the summer due to the significantly larger amount of energy produced during this time.

The daily profile generation from the three configurations is remarkably different, as shown in Fig \ref{fig_electricity_yield}(c). In Denmark, due to the price variation of electricity throughout the day, there is a dip in the electricity price during midday, as depicted in Fig. \ref{fig_electricity_yield}(c). The vertical bifacial generation profile matches better the electricity price profile. This means that although the single-axis tracking has higher electricity yield, the vertical bifacial shows higher price-weighted electricity yield. This is also true for all spacing and height pairings in Foulum but might be different in other locations.

\begin{figure*}[!h]
\centering
\begin{subfigure}{0.3\textwidth}
\includegraphics[width=\linewidth]{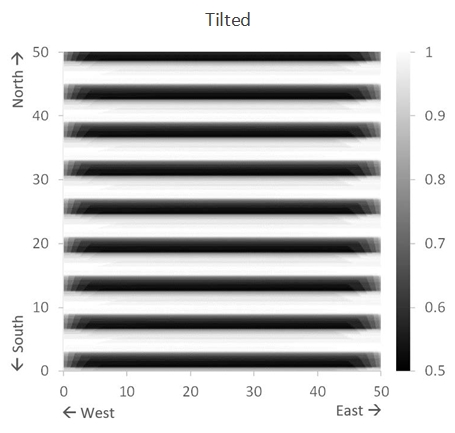}
\caption{}
\label{fig:subim1}
\end{subfigure}
\begin{subfigure}{0.3\textwidth}
\includegraphics[width=\linewidth]{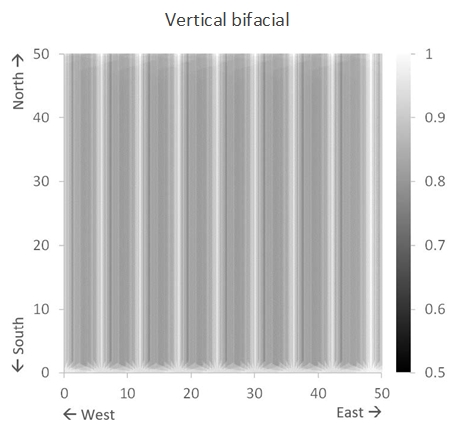}
\caption{}
\label{fig:subim1}
\end{subfigure}
\begin{subfigure}{0.3\textwidth}
\includegraphics[width=\linewidth]{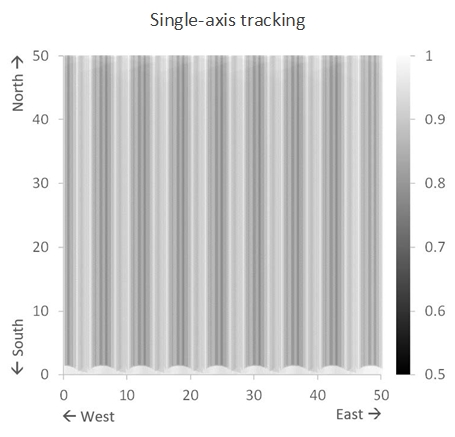}
\caption{}
\label{fig:subim1}  
\end{subfigure}
\begin{subfigure}{0.8\textwidth}
\includegraphics[width=\linewidth]{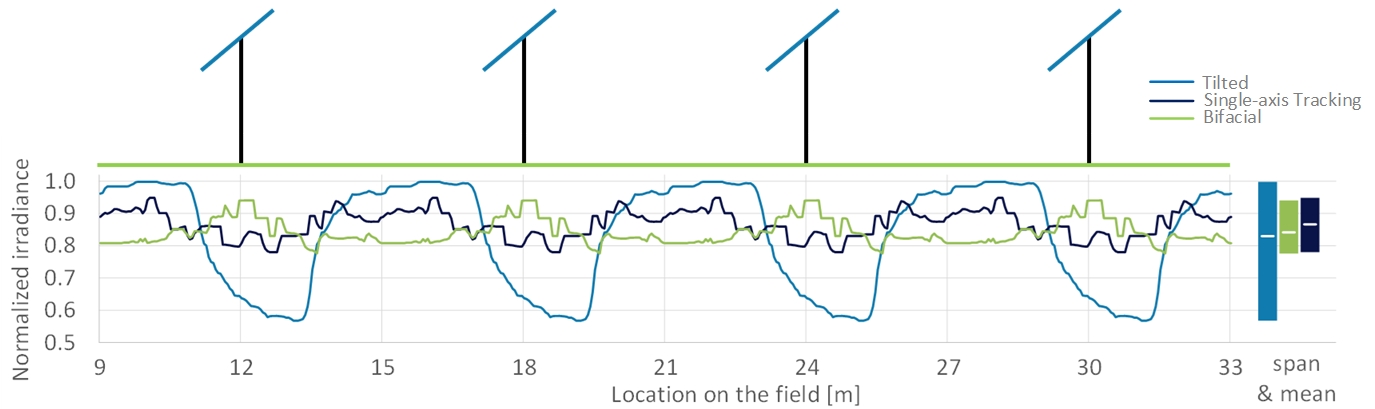}
\caption{}
\label{fig:subim1}
\end{subfigure}

\caption{ Integration of ground shadowing throughout July for the reference location (Foulum, latitude: 56.49$^{\circ}$, longitude: 9.57$^{\circ}$) for: (a)  static with optimal tilt, (b) vertical mounting, (c) single-axis horizontal tracking, and (d) line graph showing the normalized irradiance for all three configurations in July (reference setup: height= 2m, spacing= 6m).} \label{fig_heatmaps} 
\end{figure*}

Let us turn now to the analysis of the irradiance reaching the ground. The heat maps in Fig. \ref{fig_heatmaps} show the irradiance distribution on the ground for all three setup types using the reference installation mentioned above. The pixels are colored according to the percentage of total irradiance reaching the field in a grayscale where white is used for the parts of the field that do not experience shadowing.

The irradiance distribution in July in the case of the tilted setup in Foulum is shown in Fig. \ref{fig_heatmaps}(a). It is visible that some areas of the field experience significantly more shading than others. Each of the south-facing rows of PV panels casts a strip of shadow on the ground behind and under it. Here only about 56\% of incident radiation reaches the ground. This uneven and very distinct irradiance distribution could cause irregularities in plant growth.

The vertical bifacial configuration, whose irradiance distribution is plotted in Fig. \ref{fig_heatmaps}(b) depicts that the shadows and the irradiance reaching the ground are evenly distributed over the field. While there are sill strips with a lower incident irradiance along the entire length of the field, the minimal irradiance reaches 77.6\% of available irradiance (i.e., without shadowing). 

Furthermore, the single axis-tracking configuration results in a fairly even irradiance distribution over the field (see Fig. \ref{fig_heatmaps}(c)). The strips with the highest amount of shadow are still illuminated by 78\% of available irradiance. The general shadow distribution is very similar to the vertical bifacial setup but the location is shifted. For this setup, the shaded area is underneath the row of solar panels, while the area in between the panels experiences the highest irradiance. This is more convenient because the area beneath the solar panel structure is more difficult to be accessed by agricultural machinery.  

The three setups have a similar average normalized irradiance between 82\% and 86\%, which shows that the mean value does not provide sufficient information about the differences in irradiance patterns. Therefore, it is not possible to conclude plant growth potential from it. In continuation, the graphs in Fig. \ref{fig_heatmaps}(d) clearly show the differences between the setups. The tilted setup has the biggest span with values between 57\% and 100\% of available irradiance. In contrast to that, the range of the vertical bifacial and axis-tracking setup is narrower, spanning from 78\% to 94\%. 

\begin{figure}[!h]
\centering
\includegraphics[width=\linewidth]{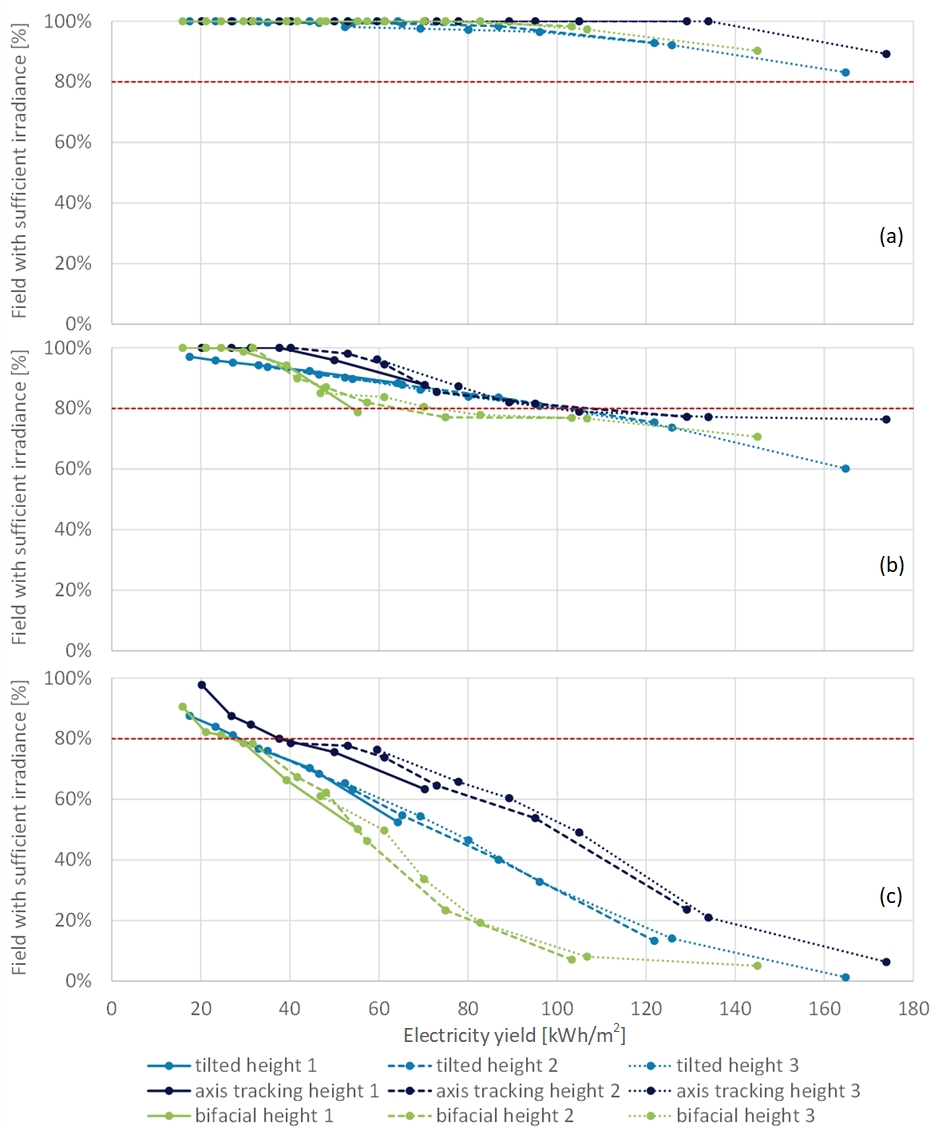}
\caption{Percentage of land useful for crops plotted as a function of the annual electricity yield (kWh/m$^2$) for the three configuration described in Fig. \ref{configurations} for Foulum. Field percentage eligible for (a) low/radiation, (b) medium/radiation, and (c) high/radiation demand crops.} \label{fig_decission_map} 
\end{figure}

If we combine now the estimation of irradiance reaching the ground and crop sensitivity (see Fig. \ref{fig_crops_model}). Fig. \ref{fig_decission_map} shows the fraction of the field with sufficient irradiance for the crops, as a function of the electricity yield. Hence, it can be seen as a decision map where either the crop production (left upper corner) or the electricity production (right lower corner) can be maximized, or a trade-off between both can also be selected. For crops with a low or mid-radiation threshold, high electricity yields can be achieved while maintaining more than 80\% of the land suitable for crops. For high-radiation-demand crops, high electricity yields (obtained by high capacity density for the PV panels) cause a significant drop in the land suitable for crops. The single axis-tracking enables a higher combination of electricity yield and land availability for crops, but at the expense of higher costs. 

We set up a target of maintaining at least 80\% of the land suitable for crops. In this case, for the high-radiation-demand crops, the annual electricity yield for the tilt and bifacial vertical configurations is similar and limited to 30 kWh/m$^2$ (see Fig. \ref{fig_decission_map}). This corresponds to a capacity density of around 30 W/m$^2$ and this is the reference value that we will use the estimate the potential for APV in different European regions in coming sections. 
% \ref{sec_APV_potential}.

\FloatBarrier
\subsection{Extension of the analysis to Europe}

As expected, the amount of electricity produced annually by every PV installation, that is, the specific yield, increases for all three configurations as the latitude decreases. The specific yield for the axis-tracking setup is the highest in every location, followed by the tilted setup and lastly the vertical bifacial setup.

\begin{figure}[!]
\centering
\includegraphics[width=\linewidth]{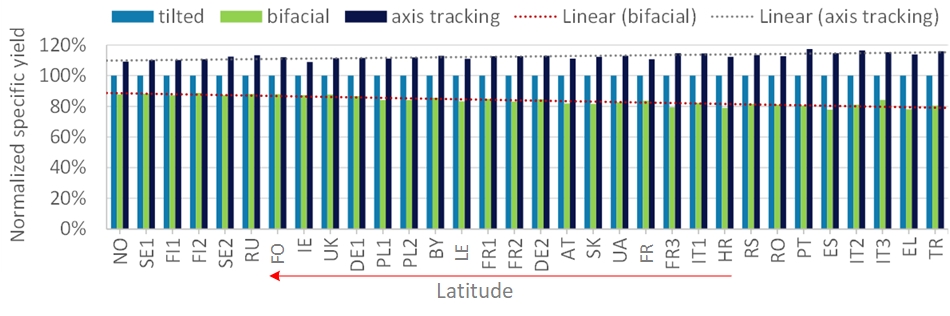}
\caption{ Normalized specific yield at the different locations (normalized by tilted setup).} \label{Normalized_specific_yield} 
\end{figure}

The order of three different setups is the same, regardless of the location. In addition to this, it is also investigated whether the relative difference between setups is also independent of the location. To investigate this, the specific yield for the different setups is normalized using as a reference the tilted installations in the respective location, as depicted in Fig. \ref{Normalized_specific_yield}.

In that figure, the locations are arranged by decreasing latitude from left to right. The differences are more notable as latitude decreases. In other words, the gain in electricity yield by tracking is less important in Northern European countries and so are the losses in electricity yield of vertical bifacial compared to the tilt configuration. There are some countries for which the values deviate from this pattern, which is due to local weather influences.

\begin{figure}[!h]
\centering
\includegraphics[width=\linewidth]{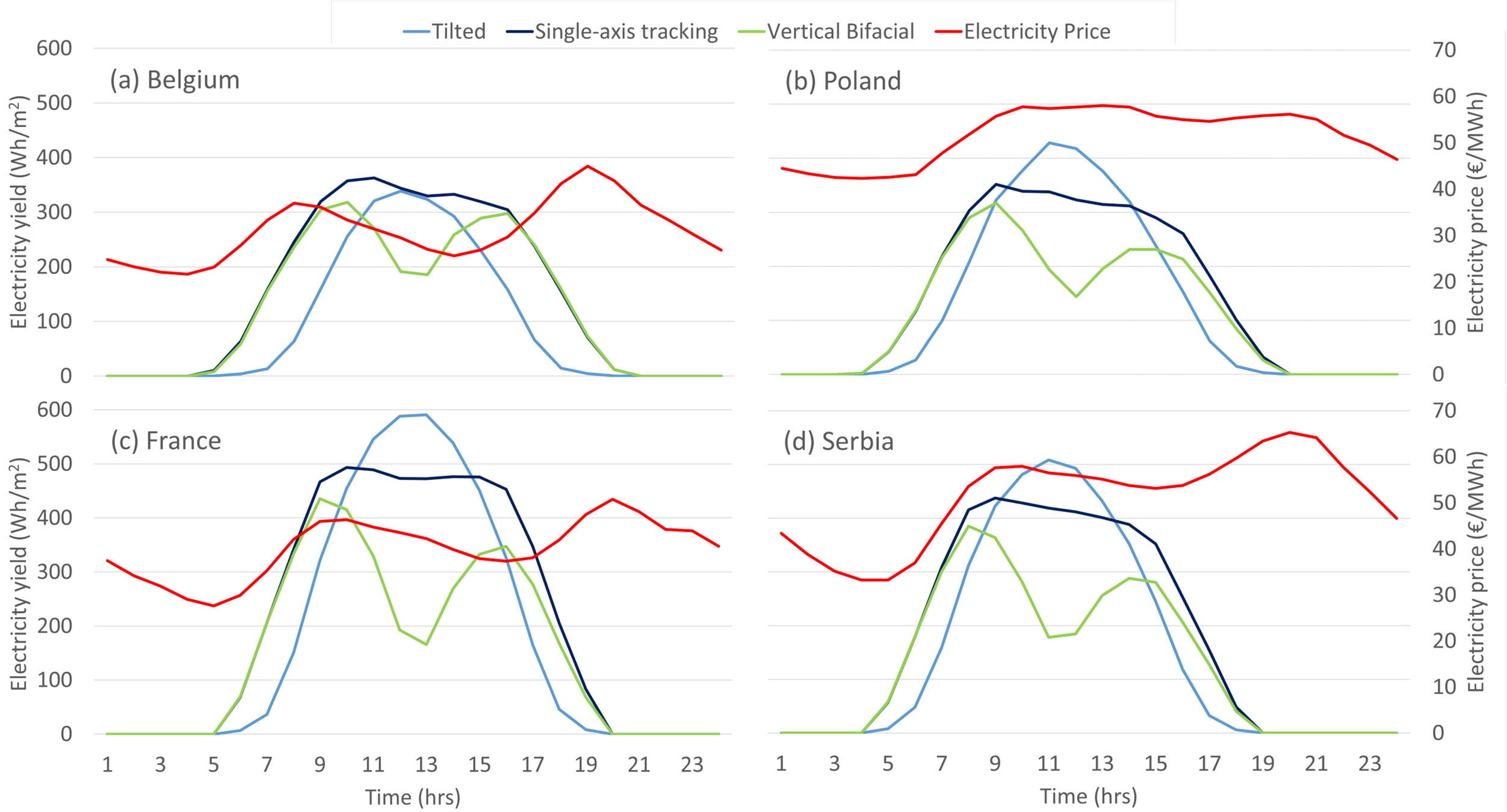}
\caption{ Electricity generation and electricity price throughout the day for the locations: (a) Belgium, (b) Poland, (c) France, and (d) Serbia.} \label{Electricity_generation_price} 
\end{figure}

Figure \ref{Electricity_generation_price} extends Fig. \ref{fig_electricity_yield} to other countries in Europe. The latitude slightly modifies the daily generation profile for the three different configurations. More notably, the daily evolution of spot market electricity prices is different in every country since they are mainly dictated by the demand pattern and renewable penetration. Overall, the tracking configuration always attains the highest electricity yield. However, the tilted or the bifacial vertical configuration achieves the highest price-weighted electricity yield (Eq \ref{eqn_price_weighted_yield}), depending on the specific country, as shown in Fig. 9.

\begin{figure}[!h]
\centering
\includegraphics[width=\linewidth]{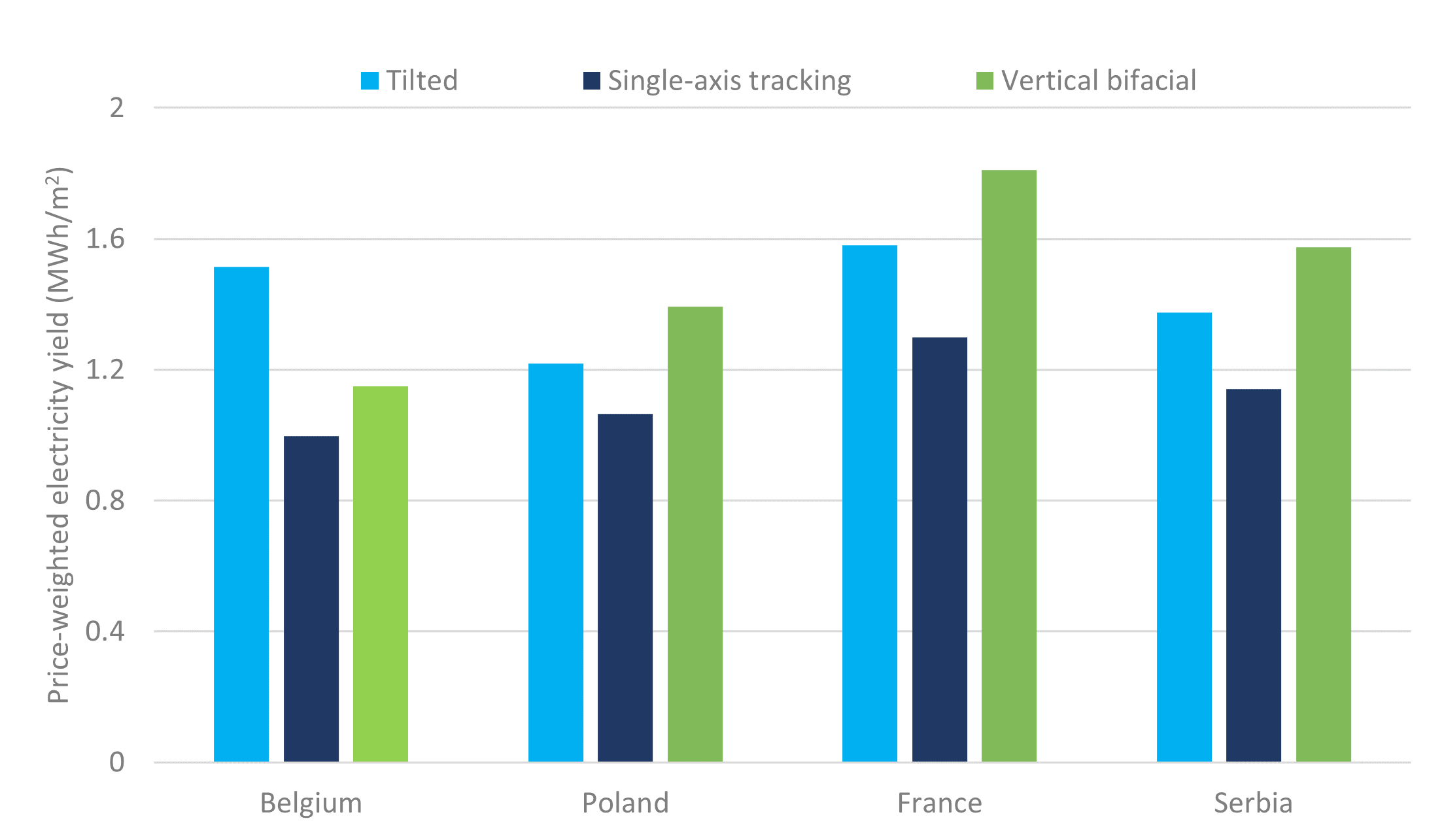}
\caption{ Price-weighted electricity yield for Belgium, Poland, France, and Serbia.} \label{Price_weighted_EY} 
\end{figure}

\subsection{Eligibility area and potential for APV for different European regions}
To assess the potential of APV in Europe for every NUTS-2 region, the feasibility and the extent to which it is possible to generate electricity are investigated in this section. To determine, which areas within a country are suitable for the use of APV, a land eligibility analysis is performed.

\begin{figure}[!]
\centering
\includegraphics[width=0.458\textwidth]{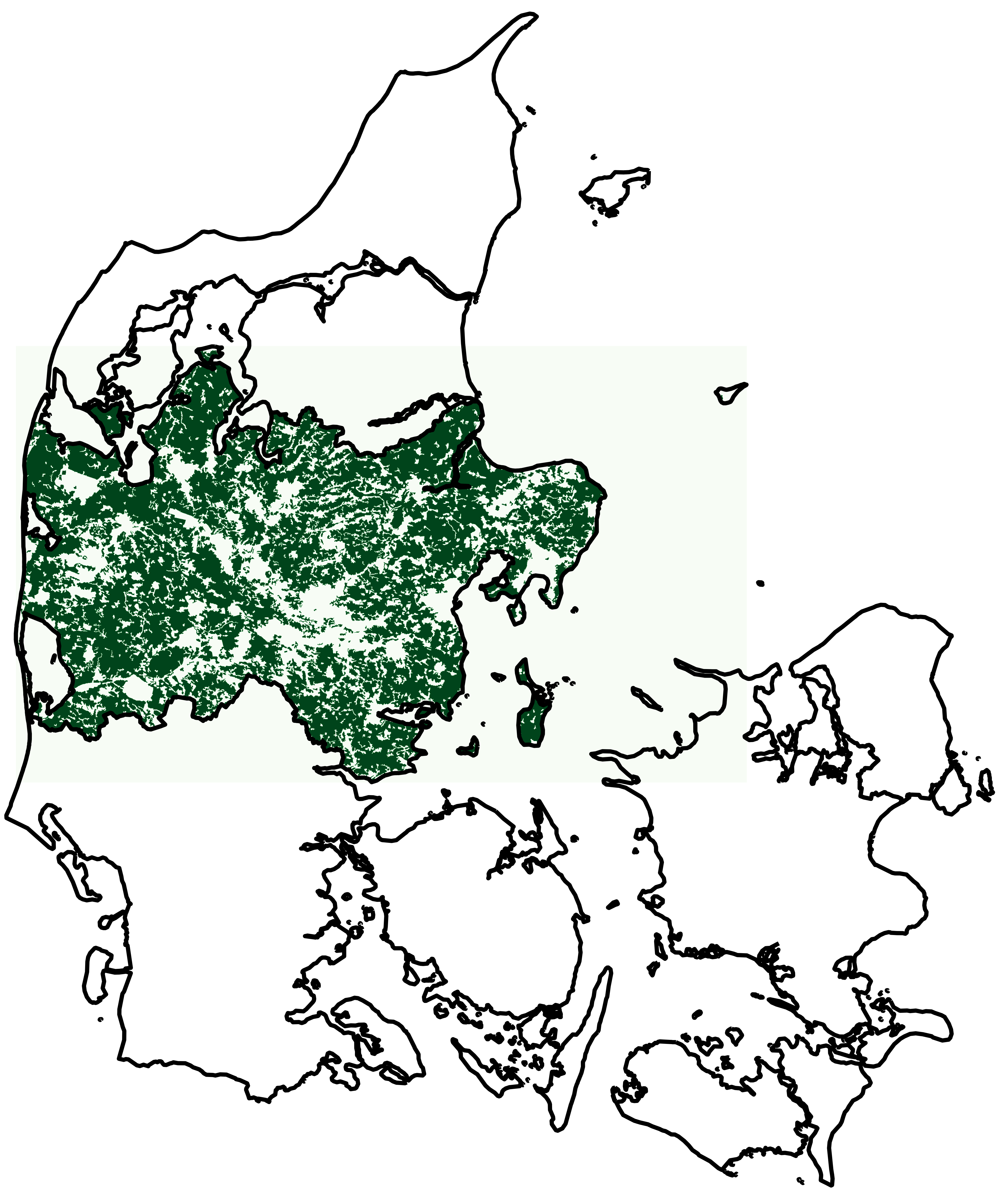}
\caption{ Suitable eligible land for APV installations in Midtjylland (Denmark). The eligible area (8341km$^2$) represents 64\% of the total area of Midtjylland.} \label{fig_eMidtjylland} 
\end{figure}

There are various land types, e.g., artificial surfaces, agricultural areas, forest and semi-natural areas, wetlands, and water bodies. The agricultural areas that are most suitable for APVs include arable land, permanent crops, and pastures (see Table 1). In these selected areas, the eligible area for APVs in the central Denmark region, i.e., Midtjylland is around 64\%, and the total eligible area is 8341 km$^2$, which is shown in Fig. \ref{fig_eMidtjylland}. Assuming a capacity density of 30 W/m$^2$, obtained in Section 4.1, and specific yield of 850 kWh/kW corresponding to vertical bifacial in Foulum, the potential electricity generation from APV installations in Midtjylland represents 215 TWh/year. Additionally, the potential estimation for every NUT-2 regions in Europe is provided in Zenodo: 7267022.

\begin{figure}[!h]
\centering
\includegraphics[width=0.458\textwidth]{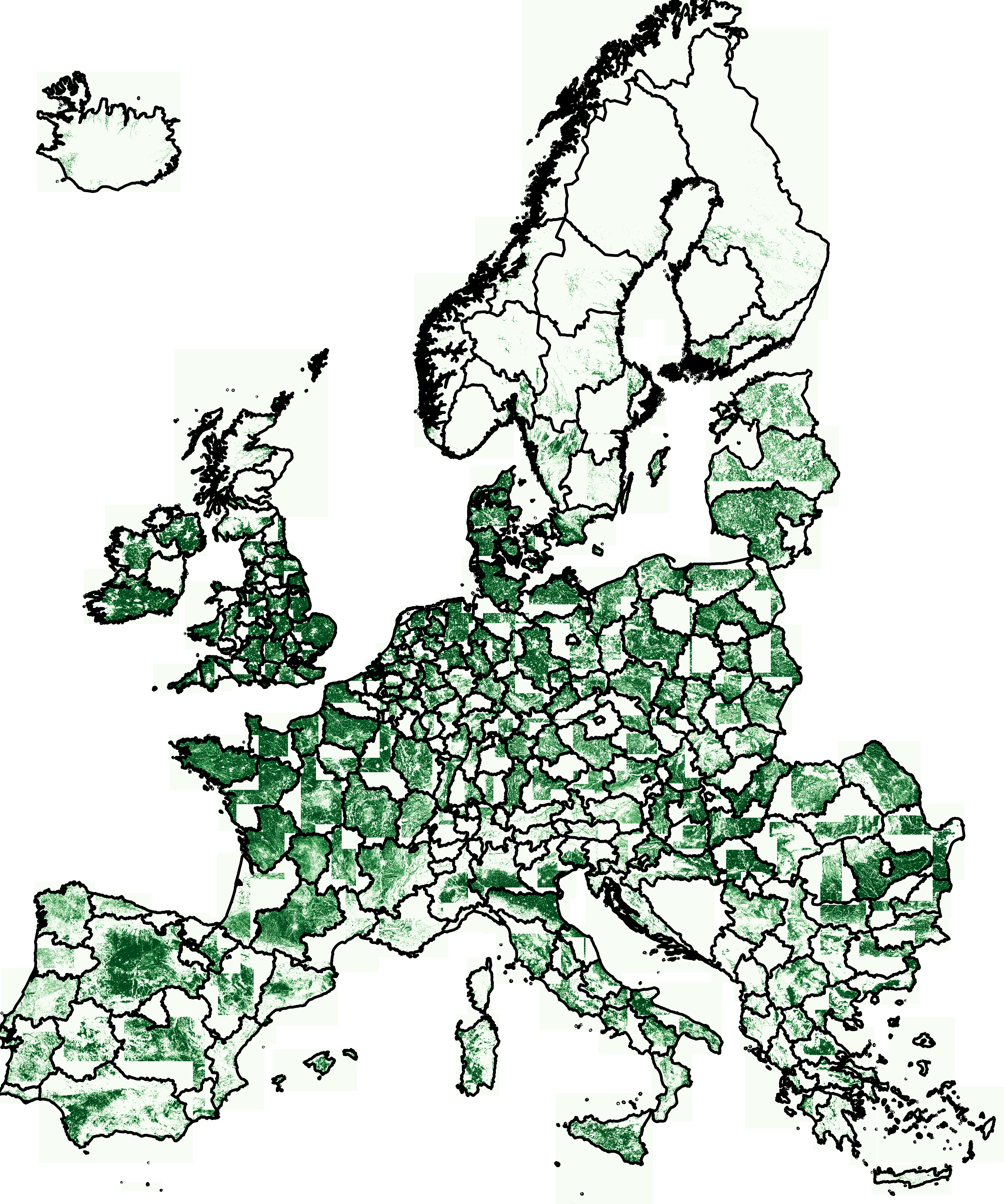}
\caption{Suitable eligible land for APV installation in the NUTS-2 regions. The eligible area represents 16.2\% area of EU or 1.7 million km$^2$.} \label{eligibility_europe} \end{figure}

Similarly, the land eligibility analysis for APV is extended to Europe considering the NUT-2 regions, and the eligible regions for APVs in Europe are shown in Fig. \ref{eligibility_europe}, which represent 16.2\% and the total eligible area is 1.7 million km$^2$. The eligible area is quite unevenly distributed across Europe. For the majority of the investigated countries in Europe, the share of eligible land is between 12\% and 29\%. Few countries have an even lower percentage with 1\% to 9\%. Several countries have higher percentage of available land, and these countries are Hungary with 58.6\%, Denmark with 53.9\%, and Ireland with 63.9\% of eligible APV area. Land area currently used for fruit trees can be of particular interest for static tilt APV installations because they can protect the trees from heavy rainfall or hail. The available land in Europe for fruit trees corresponds to approximately 29000 km$^2$.

\begin{figure}[!]
\centering
\includegraphics[width=\linewidth]{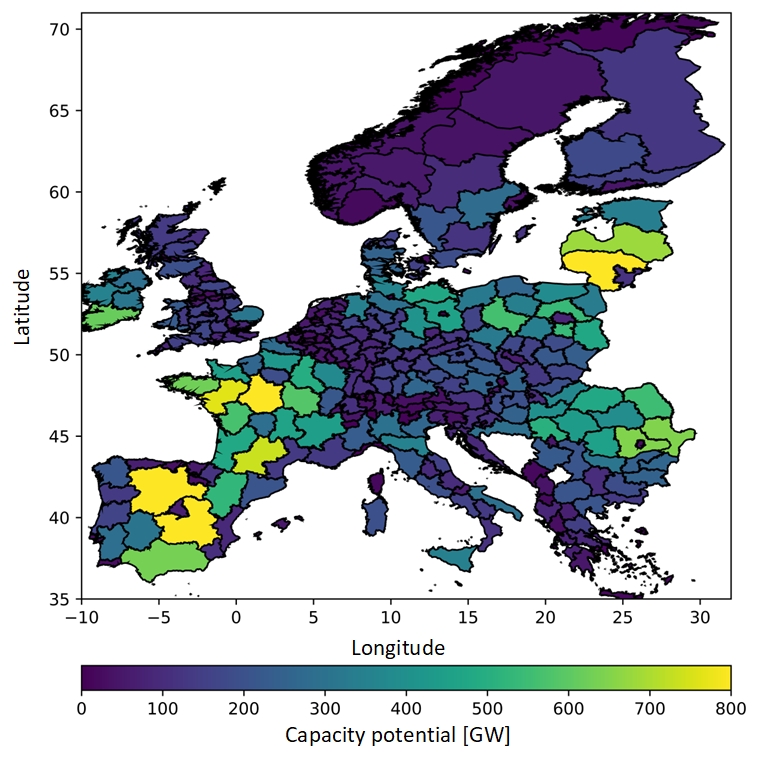}
\caption{ Maximum capacity potential for APV systems estimated for every NUTS-2 region based on the land availability in Fig. \ref{eligibility_europe} and assuming a capacity density of 30W/m$^2$ for the selected area in Table 1.} \label{Fig_capacity_potential}
\end{figure}

Furthermore, the capacity potential (in GW) in the NUTS-2 regions for APV is found by considering the land types indicated in Table 1 and a capacity density of 30 W/m$^2$. The result in Fig. \ref{Fig_capacity_potential} shows that the southern and eastern parts of Europe are more suitable for APV systems, which agrees with the results in \cite{willockx2022geospatial}. 

\begin{figure}[!h]
\centering
\includegraphics[width=0.45\textwidth]{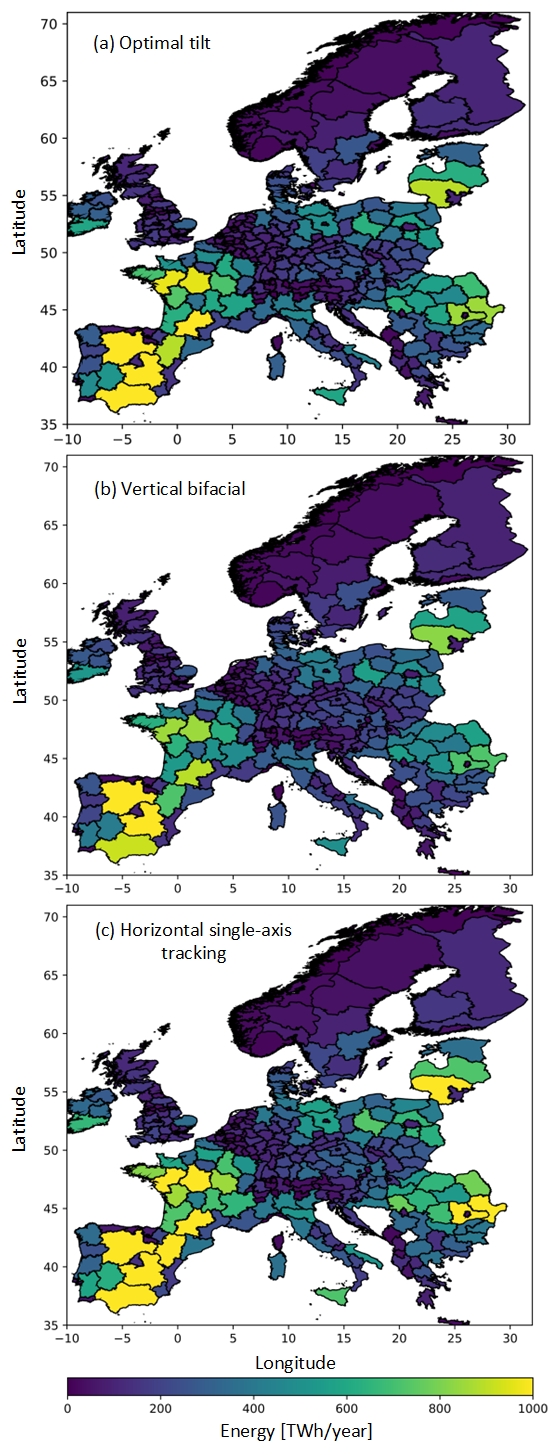}
\caption{Annual energy production for different NUT-2 regions by considering a capacity density of 30W/m$^2$ and agricultural land types defined in Table 1: (a) optimal tilted, (b) vertical bifacial, and (c) horizontal single-axis tracking.} \label{Fig_energy}
\end{figure}

The potential energy production (in TWh/year) for the NUTS-2 regions in the EU for different APV systems is shown in Fig. \ref{Fig_energy} for optimal tilt, vertical bifacial, and horizontal single-axis tracking PV systems. The energy production over the year is found by using irradiance data from PVGIS \cite{PVGIS} corresponding to the central coordinates of every NUTS-2 region. The enormous potential of APV electricity generation is note worthy. The vertical bifacial APV systems can produce up to 71,500 TWh per year in the EU, which is 28 times higher than the current electricity demand. In some countries like Denmark, energy production can reach up to 26 times the current production.

% \begin{figure}[!]
% \centering
% \begin{subfigure}{0.38\textwidth}
% \includegraphics[width=\linewidth]{energy_fixtilt.png}
% \caption{}
% \label{fig:subim1}
% \end{subfigure}
% \begin{subfigure}{0.38\textwidth}
% \includegraphics[width=\linewidth]{energy_bifacial.png}
% \caption{}
% \label{fig:subim1}
% \end{subfigure}
% \begin{subfigure}{0.38\textwidth}
% \includegraphics[width=\linewidth]{energy_tracking.png}
% \caption{}
% \label{fig:subim1}
% \end{subfigure}
% %\includegraphics[width=\textwidth]{yield_static_vertical.png}
% \caption{ Annual energy  production for different NUT-2 regions by considering a capacity density of 30W/m$^2$ and agricultural land types defined in Table 1: (a) optimal tilted, (b) vertical bifacial, and (c) horizontal single-axis tracking.} \label{Fig_energy}
% \end{figure}

%%%%%%%%%%%%%%%%%%%%%%%%%%%%%%%%%%%%%%%%%%%%%%%%%%%%%%%%%%%%%%%%%%%%%%%%%%%%%%%%%%%%%%%%%%%%%%%%%%%

\section{Conclusion} % Marta (500 words)
This work investigates different types of agrivoltaic (APV) configurations in Foulum, Denmark along with other locations in Europe based on NUTS-2 regions. Two different characteristics are used to study the feasibility of APV systems, i.e., the potential of PV systems, and their influence on the underlying farmland. During the study, three different APVs setups are considered, i.e., optimal tilted, horizontal single-axis tracking, and vertical bifacial.

In this work, a model was developed, which simulates the shadows on solar panels and the ground. The model allows the accurate analysis of the reduced production output due to shadow losses for each simulated hour, rather than just assuming a general loss factor. This is an important factor when comparing and analyzing the three different APVs setups mentioned above.

As expected, the axis-tracking setup produces a higher electricity yield, but when taking into account the daily generation patterns of the different configurations, the vertical bifacial produces a higher price-weighted electricity yield.

A capacity density of around 30 W/m$^2$ is used to estimate the potential for APV in different NUT-2 regions in Europe because it helped in achieving high electricity yields along with maintaining the target of keeping more than 80\% of the land suitable for crops.

Furthermore, the eligible areas for APVs in Europe are determined using the Corine Land Cover database, and applying constraints like distance to forests, settlements, and roads, while ensuring that the area is on land that already is used for agriculture. This analysis shows that the eligible area is distributed quite unevenly across Europe, with some countries (e.g., Norway) having as little as 1\% of their total area suitable for APVs, while in others this percentage is as high as 53\% (e.g., Denmark). Overall, APV has great potential with a potential capacity of 51 TW in Europe and can produce up to 71500 TWh per year, which is 28 times high than the current electricty demand in Europe.

\section{Acknowledgements}

The authors would like to express their acknowledgment to Ann-Sophie Reimer and Lars Rodrigo Wagner for their master’s thesis at our research group which produced several of the results in this paper. We would also like to thank Uffe Jørgensen, Johannes Wilhelmus Maria Pullens, Carl-Otto Ottosen, and Gabriele Torma for fruitful discussions around this paper. The authors are fully or partially funded by the HyPErFarm (Hydrogen and Photovoltaic Electrification on Farm) project, which is supported by the EU2020 under grant number 101000828. The responsibility for the contents lies solely with the authors.

\bibliography{agrivoltaics}

\end{document}